\documentclass[twocolumn,superscriptaddress,amsmath,amssymb,aps,prb]{revtex4-1}
\usepackage{adjustbox}
\usepackage[hidelinks]{hyperref}
\usepackage{graphicx}
\usepackage{dcolumn}
\usepackage{bm}

\begin{document}

\title{Evolution of Nagaoka phase with kinetic energy frustrating hoppings}

\author{F. T. Lisandrini}
\affiliation{Instituto de F\'{\i}sica Rosario (CONICET) and Universidad Nacional de Rosario,
Bvd. 27 de Febrero 210 bis, (2000) Rosario, Argentina}
\author{B. Bravo}
\affiliation{Instituto de F\'{\i}sica de La Plata, UNLP-CONICET and Departamento de F\'{\i}sica, Facultad de Ciencias Exactas, Universidad Nacional de La Plata, 1900 La Plata, Argentina.}
\author{A. E. Trumper}
\author{L. O. Manuel}
\author{C. J. Gazza}
\affiliation{Instituto de F\'{\i}sica Rosario (CONICET) and Universidad Nacional de Rosario,
Bvd. 27 de Febrero 210 bis, (2000) Rosario, Argentina}

\date{\today}

\begin{abstract}
We investigate, using the density matrix renormalization group, the evolution of the Nagaoka state with 
$t'$ hoppings that frustrate the hole kinetic energy in the $U=\infty$ Hubbard model on the anisotropic 
triangular lattice and the square lattice with second-nearest neighbor hoppings. We find that the 
Nagaoka ferromagnet survives up to a rather small $t'_c/t \sim 0.2.$ At this critical value, there is a 
transition to an antiferromagnetic phase, that depends on the lattice: a ${\bf Q}=(Q,0)$ spiral order, 
that continuously evolves with $t'$, for the triangular lattice, and the usual ${\bf Q}=(\pi,\pi)$ N\'eel 
order for the square lattice. Remarkably, the local magnetization takes its classical value for all 
considered $t'$ ($t'/t \le 1$). Our results show that the recently found classical kinetic 
antiferromagnetism, a perfect counterpart of Nagaoka ferromagnetism, is a generic phenomenon in these 
kinetically frustrated electronic systems. 

 \begin{description}
\item[PACS numbers] 75.10.Lp, 71.10.Fd
 \end{description}

\end{abstract}

\maketitle

\section{Introduction} 

Nagaoka's theorem \cite{nagaoka66} stands almost alone as a rigorous result about itinerant magnetism. 
It predicts the existence of a fully polarized ferromagnetic state as the unique ground state of 
the $U=\infty$ Hubbard model, when one hole is doped away half-filling and certain connectivity 
conditions are satisfied. Despite its very restricted validity and its thermodynamic irrelevance, 
the theorem introduced an interesting idea about quantum magnetism: kinetic magnetism, the possibility 
of magnetic order driven solely by the motion of the electrons. 

Since the seminal work by Nagaoka \cite{nagaoka66}, a lot of effort has been dedicated to the study 
of Nagaoka ferromagnetism stability beyond the constraints of the theorem. In particular, some 
controversy arose as to whether the fully polarized state survives for a finite density of holes
(see \cite{hartmann93, fazekas99,park08} and references therein). However, large-scale density 
matrix renormalization group (DMRG) calculations \cite{liu12}, among others \cite{park08,liang95,carleo11}
seem to have solved the problem, unless for the square lattice, as they predict the existence of Nagaoka 
ferromagnetism up to critical hole density $\delta_c \simeq 0.2$. Little is known about the 
states that supplant the Nagaoka ferromagnet beyond $\delta_c$ \cite{liu12,carleo11}. 

With respect to the $U=\infty$ condition, its relaxation leads to the competition between the 
Nagaoka and antiferromagnetic exchange mechanisms. This entails the instability of the Nagaoka phase 
against phase separation:  for $U/t \lesssim 130$, a ferromagnetic polaron around the hole  
moves on an antiferromagnetic background \cite{eisenberg02}. 

Lastly, the violation of the connectivity condition \cite{oles91,park08} can also destabilize the 
Nagaoka phase. Nagaoka's theorem requires that $\mathbb{S}_{loop}=1,$ where 
$\mathbb{S}_{loop}$ is the sign of the hopping amplitudes around the smallest loop of the lattice. 
When this condition is not fulfilled the hole kinetic energy is frustrated. Kinetic frustration 
\cite{barford91,merino06,wang08} is a quantum mechanical phenomenon, without classical analog 
since it originates in the quantum interferences of different hole paths. A simple way to break 
the connectivity condition is the consideration of particular signs for the hopping parameters in 
non-bipartite lattices. As an alternative way, the hopping integrals can be modulated by a 
staggered magnetic flux \cite{wang08}.

In 2005, Haerter and Shastry \cite{haerter05} found that the ground state of the $U=\infty$ triangular 
Hubbard with $t >0,$ a kinetically frustrated system where Nagaoka's theorem is not valid 
($\mathbb{S}_{loop}=-1$), has a $120^{\circ}$ N\'eel order. More recently, we found another example of 
kinetic antiferromagnetism, a $(\pi,\pi)$ N\'eel order as the ground state of the square Hubbard model 
with second-nearest neighbor hopping $t'=t > 0$, and we uncovered the classical nature of these 
antiferromagnets \cite{sposetti14}.  At the same time, we proposed a microscopic mechanism responsible 
for this kinetic antiferromagnetism, based on the relaxation of the kinetic frustration as the hole moves 
on an antiferromagnetic background.

In order to deepen our understanding of kinetic magnetism, in this work we study the evolution of the 
Nagaoka ferromagnet with kinetically frustrating $t'$ hoppings. To make it, we solve the $U=\infty$ Hubbard 
model on the anisotropic triangular lattice and the square lattice with second-nearest neighbor hopping
(see Fig. \ref{redes}), using the numerically exact DMRG. As we vary $t'$ we can move between the two known 
limits: the Nagaoka state ($t'=0$) and the novel kinetic antiferromagnet ($t'=t$) \cite{sposetti14}. 
We find that the classical order extends for all $t'$, ferromagnetic below a critical $t'_c$ (that slightly
depends on the lattice), and antiferromagnetic above. We analyze the characteristics of the transition, the 
antiferromagnetic structure above $t'_c$, and the physical microscopic mechanism at work for each lattice. 

Beyond the sustained theoretical interest in kinetic magnetism for decades, up to date, there are no 
clear experimental realization of Nagaoka conditions. At the end of this work, we briefly mention some 
recent experimental proposals.

\section{Model and method}
\label{model}
\begin{figure}[ht]
\centering
\includegraphics[width=0.4\textwidth]{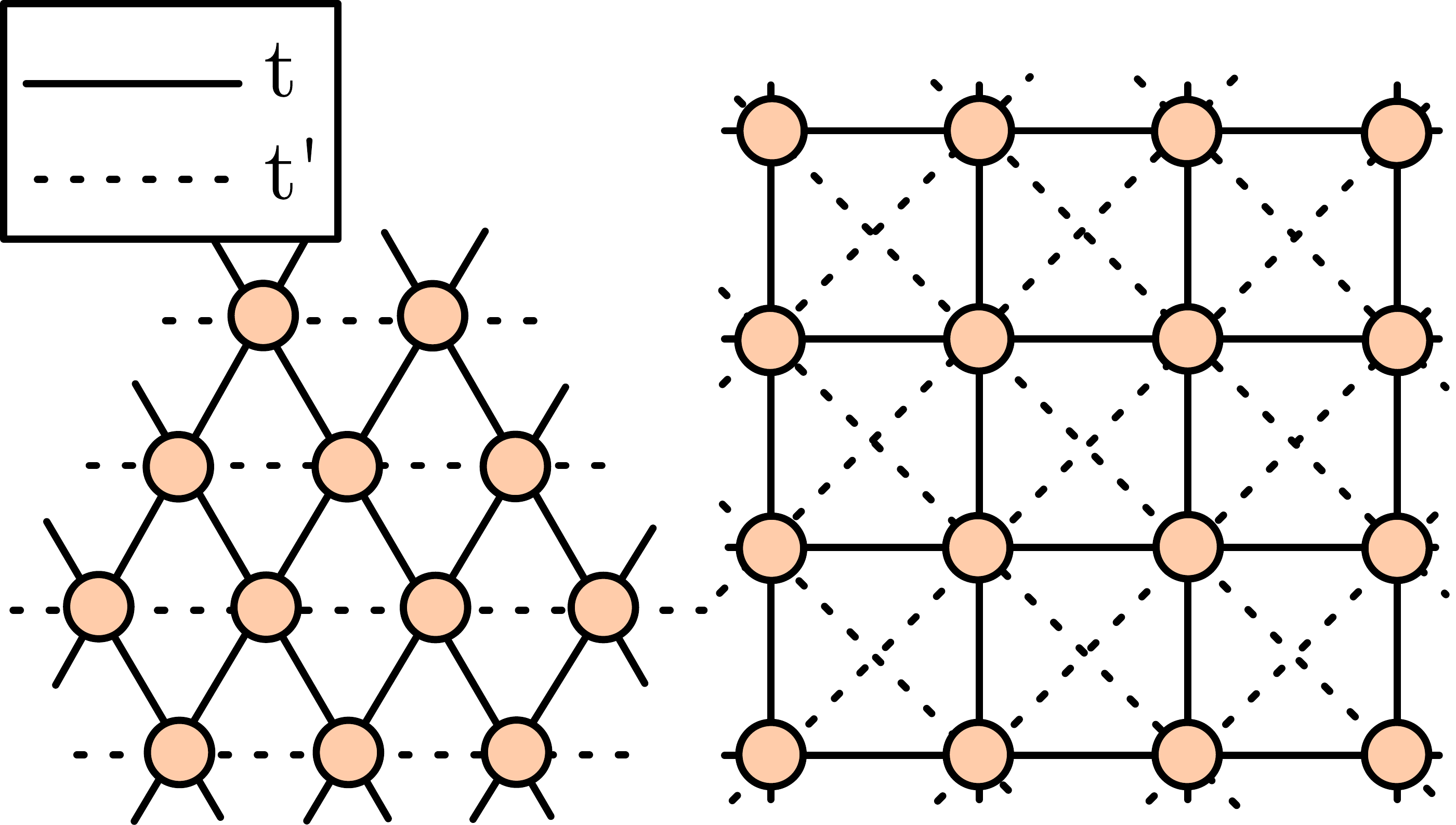}
\caption{Triangular lattice with spatially anisotropic hoppings and square lattice 
with second-nearest neighbor hoppings. $t's$ are the kinetically frustrating hoppings. 
We take $t=1$ as the energy unit.}
\label{redes}
\end{figure}

In order to analyze the stability of the Nagaoka ferromagnetic state against kinetic 
frustration, we study the $U = \infty$ Hubbard model, with one hole doped away the 
half-filled case, on two lattices: the spatially anisotropic triangular lattice and 
the square lattice with second-nearest neighbor hoppings, as shown in Fig. \ref{redes}.  
The Hubbard model reads
\begin{equation}
 \label{Hubbard}
\hat{H}=-\sum_{ij\sigma}t_{ij}\hat{c}^{\dagger}_{i \sigma}\hat{c}_{j\sigma}+
U\sum_{i} \hat{n}_{i\uparrow}\hat{n}_{i\downarrow},
\end{equation}
where $i,j$ denote pair of sites on each lattice, $t_{ij}$ are the hopping integrals, 
and $U$ is the onsite Coulomb repulsion. In Fig. \ref{redes}, the solid lines represent 
$t$ (which we take as the energy scale, $t=1$), while the dashed lines represent $t'$, the 
varying anisotropic hoppings (second-nearest neighbor hoppings) for the triangular 
(square) lattice. It should be noticed that, for $t' = 0,$ the Hubbard model on the 
two lattices are equivalent. 
When $t' = 0$, the connectivity condition of Nagaoka's theorem is fulfilled because 
the minimal loops for the hole motion are squares with 
$\mathbb{S}_{loop}= {\text sgn}(-t)^4 = +1$. For finite $t'$, the 
minimal loops are triangles with 
$\mathbb{S}_{loop} = {\text sgn}(-t')(-t)^2 = -{\text sgn (t')}.$ So, for $t' \le 0$, 
the theorem is valid and the ground state is a unique fully polarized 
ferromagnetic state for both lattices. On the other hand, a positive $t'$ introduces 
kinetic energy frustration in the hole motion ($\mathbb{S}_{loop}=-1$), invalidating Nagaoka's 
theorem. In a previous work \cite{sposetti14}, we have shown that, in the 
special case $t'=t$, the ground states have classical antiferromagnetic orders:  a 
$120^\circ$ pattern for the (isotropic) triangular lattice and the usual 
${\bf Q}= (\pi,\pi)$ for the square one. 

In this work we will use DMRG \cite{white92,white93} to solve the ground state of the
$U=\infty$ Hubbard model, for $0 \le t' \le  1$. We apply DMRG to ladders of 
dimension $L_x \times L_y$, 
with $L_y=6$ legs (enough to describe correctly two-dimensional systems \cite{liu12}) 
and up to $L_x=15$ rungs. We impose cylindrical boundary conditions with periodic wrapping in 
the rung direction and open boundary conditions along the legs. To maintain  errors of 
the DMRG smaller than symbol sizes in each figure, we have kept up to $m=500$  states, 
with a truncation error less than $O(10^{-7})$.

\section{Results}

\subsection{Ground state energy and critical $t'_c$}

As we mentioned above, the $t'=0$ ground state of Hamiltonian (\ref{Hubbard}) is ferromagnetic, 
while for $t'=t$ it exhibits antiferromagnetic order for each lattice \cite{sposetti14}. 
Hence, there will be a critical value $t'_c$ where the Nagaoka state is destabilized. 
To determine $t'_c$, we have resorted to an energy analysis. Let $E_N(S^z)$ be the ground state 
energy of the $U=\infty$ Hubbard model, for an $N$-site lattice (with $N-1$ electrons) and 
a given sector of the spin projection $S^z$. For a given $t'$, we have compared the 
(ground state) energies of the different spin projection sectors, from the maximal 
$S_{max}^z =\frac{N-1}{2}$ --corresponding to a full spin polarization in the $z$ direction-- 
to the minimal $S_{min}^z = \frac{1}{2}$. Notice that, due to the $SU(2)$ symmetry of the model, 
the ferromagnetic Nagaoka state is $2S_{max}+1$-fold degenerate, where $S_{max}=\frac{N-1}{2}$ 
is its total spin, and it has projections in all the $S^z$ sectors.  
  
\begin{figure}[hb]
\centering
\includegraphics[width=0.45\textwidth]{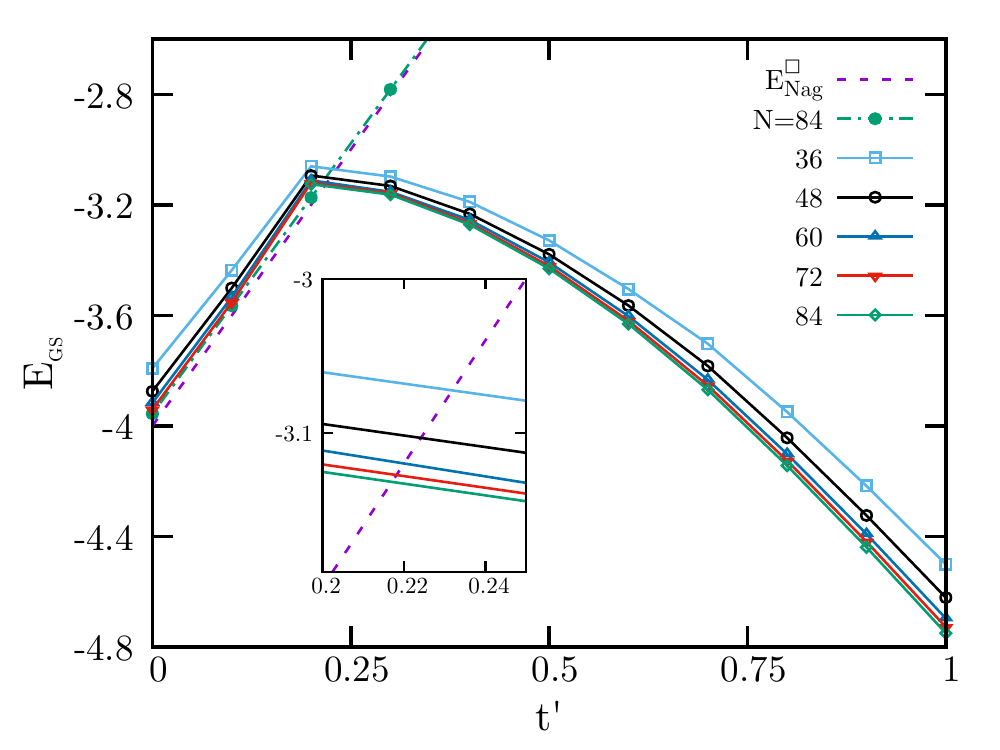}
\caption{Ground state energy $E_N(S_z)$ for the square lattice as a function of $t'$, 
and for different $N$-site clusters. 
Dashed (solid) lines correspond to the $S_z =S^{max}_z$ ($S_z = \frac{1}{2}$) 
sector energy. $E^{\square}_{Nag}$ is the thermodynamic limit of the Nagaoka state energy.
Inset: zoom of the critical region, showing the intersection of 
$E_N\left(\frac{1}{2}\right)$ with the ferromagnetic energy $E^{\square}_{Nag}$.} 
\label{ene-c}
\end{figure}

We have found that, for small values of $t'$ and for both lattices, 
the computed ground state energies of all the $S^z$ sectors are degenerate (in particular, 
$E_N\left(S_{min}^z\right)=E_N\left(S_{max}^z\right)$). Therefore, we can suspect
that these degenerate states belong to the Nagaoka ground state manifold. 
As a check, we have verified that the spin correlations, for different $S^z$ sectors, 
correspond to a fully saturated ferromagnet 
($\left<{\bf S}_i\cdot {\bf S}_j\right> \simeq \frac{1}{4}$ for $i\neq j$), discarding then the possibility 
that the Nagaoka state may be degenerate with lower total spin states.   
On the other hand, for larger values of $t'$, we have obtained that the ground state always
belongs to the minimal spin projection sector $S_{min}^z$, which would correspond to a total 
spin $S=\frac{1}{2}$. We have not found partially polarized ground states for any $t'$, 
although we must warn that, very close to the transition point $t'_c,$ 
the flattening of $E_N\left(S^z\right)$ makes the numerical treatment harder and 
less precise. With this caveat, we can say that, at $t'_c,$ there will be a transition from a 
Nagaoka ferromagnet to a minimal spin state which, later on, we will characterize as an 
antiferromagnetic state. 
\begin{figure}[ht]
\centering
\includegraphics[width=0.45\textwidth]{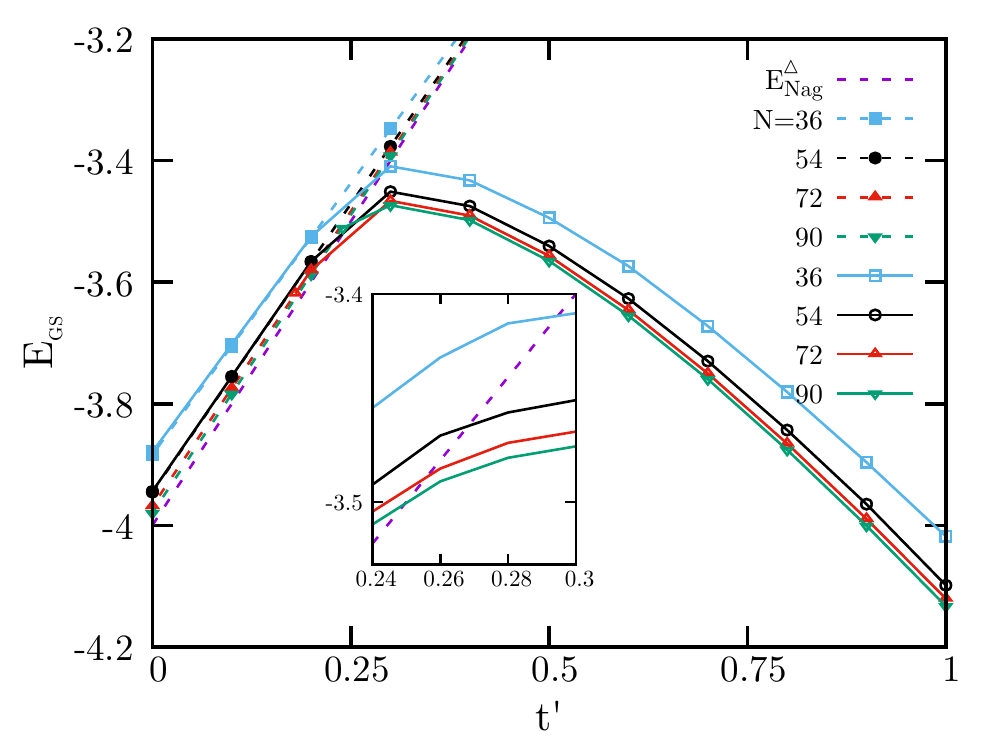}
\caption{Ground state energy $E_N(S^z)$ for the triangular lattice as a function of $t'$, 
and for different $N$-site clusters. 
Dashed (solid) lines correspond to the $S^z =S_{max}^z$ ($S^z_{min}$) 
sector energy. For smaller  $t'$ both energies overlap. 
$E^{\triangle}_{Nag}$ is the thermodynamic limit of the Nagaoka state energy.
Inset: zoom of the critical region, showing the intersection of 
$E_N\left(S^z_{min}\right)$ with the ferromagnetic energy $E^{\triangle}_{Nag}$.} 
\label{ene-t}
\end{figure}
To determine the critical $t'_c,$ firstly we have looked for the $t'$ value where the 
degeneracy between $E_N\left(S^z_{min}\right)$ and $E_N\left(S_{max}^z\right)$ is broken
for each lattice. 
In Figs. \ref{ene-t} and \ref{ene-c} 
we show these energies as a function of $t'$ and for different cluster sizes, for the 
triangular and square lattices, respectively. 
We can see that the degeneracy is broken in the region around 
$t' \sim 0.2 - 0.3$, signaling, as we explained above, that the ground state of 
the systems moves from the highest-spin Nagaoka state to another one with minimal 
total spin. 

As the conventional DMRG algorithm uses the $S^z$ quantum number without 
discriminating between different total spin $S$, we cannot compute the energy of the 
excited $S=\frac{1}{2}$ state below $t'_c$ (notice that, in this case, the calculated 
$E_N\left(S^z_{min}\right)$ corresponds to the $S^z=\frac{1}{2}$ sector energy of the 
$S_{max}$ Nagaoka ground state). Therefore, we do not have access to the expected 
energy level crossing between the highest- and lowest-spin sectors, that would 
facilitate the determination of $t'_c$. For the square lattice model, the lack of the 
level crossing is not so important as there is an appreciable kink in the ground state 
energy at $t'_c$ (see main panel of Fig. \ref{ene-c}). However, for the triangular 
case, the transition seems to be much smoother, as is shown in the main 
panel of Fig. \ref{ene-t}, and, consequently, it is more difficult to estimate the 
critical point where the degeneracy between $E_N\left(S^z_{max}\right)$ and 
$E_N\left(S^z_{min}\right)$ is lost.  To avoid this difficulty, we have evaluated 
$t'_c$ extrapolating the level crossing between $E_N\left(S^z_{min}\right)$ and the 
infinite-lattice Nagaoka energy, $E_{Nag}$ (see insets of Figs. \ref{ene-t} 
and \ref{ene-c}). $E_{Nag}$ can be computed exactly as the problem of one hole moving 
in a ferromagnetic background is identical to the spinless tight-binding 
system \cite{brinkman70}

\begin{table}[t]
\begin{center}
\begin{tabular}{lcccccc}
\hline \hline 
$N$ & $18$ &$36$ & $54$ & $72$ &  $90$ & $\infty$\\
$t'^{\triangle}_c$ & $0.387$ &  $0.295$ & $0.268$ & $0.256$ & $0.249$ & $0.222$\\
\hline \hline
\label{tab-t}
\end{tabular}
\end{center}
\caption{Critical $t'_c$ values for different $N$-site triangular clusters, and 
its $N\to \infty$ extrapolation limit.} 
\label{table_ene-t}
\end{table}

First, we consider the anisotropic triangular lattice. 
The corresponding Nagaoka ground state energy is, $E^{\triangle}_{Nag} = -4 |t| + 2t'$, 
which is shown in Fig. \ref{ene-t} with the DMRG results. 
Following the procedure mentioned above, in Table \ref{table_ene-t} we give 
the critical $t'^{\triangle}_c$ hoppings for different $N$-site clusters. We have 
extrapolated these values assuming that $t'_c \propto 1/N^2,$ and we have obtained 
$t'^{\triangle}_c \simeq 0.222$ in the thermodynamic limit.
 
\begin{table}[t]
\begin{center}

\begin{tabular}{lccccccc}
\hline \hline 
$N$ & $24$ & $36$ & $48$ & $60$ & $72$ &  $84$ & $\infty$\\
$t'^{\square}_c$  & $0.256$ & $0.232$ & $0.225$ & $0.220$ & $0.218$ & $0.217$ & $0.214$  \\
\hline \hline
\end{tabular}

\end{center}
\caption{Critical $t'_c$ for different $N$-site square clusters, and 
its $N\to \infty$ extrapolation limit.} 
\label{table_ene-c}
\end{table}
 
Second, for the square lattice with second-nearest neighbor hoppings, the exact energy of 
the fully polarized state in the thermodynamic limit is,  $ E_{Nag}^{\square} = -4 |t| + 4t'$, 
(with $t' < 0.5$) and it is shown in Fig. \ref{ene-c}. In Table \ref{table_ene-c} we present 
the critical values for different $N$-site clusters, leading to $t'^{\square}_c \simeq 0.214$ 
in the $N\rightarrow \infty$ limit. This value is quite close to the only one that existed in the 
literature up to the present, $t_c \simeq 0.255,$ obtained within a restricted Hilbert space 
\cite{oles91}. 

It is instructive to compare our results with the solution of the simplest systems that
preserve the basics of Nagaoka physics. That is, three electrons in Hubbard square 
four-site clusters with nearest-neighbor $t$ hoppings and (a) $t'$ hopping along only 
one diagonal (triangular lattice); (b) $t'$ hoppings along both diagonals 
(square lattice) \cite{park08}. In both cases, there is an energy level crossing for some $t'_c$. 
For $t' < t'_c$ the ground state has $S=\frac{3}{2}$, corresponding to the Nagaoka state, 
while for $t' > t'_c,$ the ground state has minimal spin $S=\frac{1}{2}.$  
For system (a) the transition occurs at $t'_c/t = \frac{1}{\sqrt{14}} \simeq 0.267,$ 
while for system (b) $t'_c/t = 0.25$. These values roughly correspond to the gap 
energy $\Delta$ between the $S=\frac{3}{2}$ and $S=\frac{1}{2}$ ground states for $t'=0$, that is, 
$\Delta = \left(2-\sqrt{3}\right) t \simeq 0.267 t$.
It is remarkable that the critical $t'_c$ values for these toy models are very close 
to the thermodynamic limit ones that we have presented above 
(Tables \ref{table_ene-t} and \ref{table_ene-c}). 
From this fact, we can deduce that the relevant quantum interferences for the Nagaoka physics 
are those associated with the hole motion along the smallest lattice loops. 

We want to draw attention to the fact that the critical hoppings for both lattices are 
very similar ($t'^{\triangle}_c \simeq 0.222$, $t'^{\square}_{c} \simeq 
0.214$), and also they are numerically similar to the critical doping for the destabilization 
of the Nagaoka state for $t'=0$, that is, $\delta_c \simeq 0.2$ \cite{liu12,liang95}. 
We guess that this agreement is not casual: if the Nagaoka ferromagnet, with $t'=0$ and one hole 
doped, is separated by an energy gap $\Delta$ of other spin sectors, it is plausible that
a ``perturbation'' may destabilize the phase as long as its characteristic energy is of the 
order of $\Delta$ (In the case of doping, we can associate it with an energy 
$\varepsilon \propto \delta \times t$). So, if this argument is correct, we expect a gap of 
the order of $0.2 \;t$ for the $t'=0$ Nagaoka ferromagnet in the square lattice, 
a value close to the gap for the 4-site cluster system.

\subsection{Magnetic wave vector}

The magnetic properties of the ground state can be inferred from the static magnetic structure factor 
${\cal S}(\bf k)$ defined as
\begin{equation}
{ {\cal S}({\bf k})={\frac{1}{N}} \sum_{ij}\langle {\bf S}_i.{\bf S}_j\rangle  
e^{i{\bf k}({\bf R}_i-{\bf R}_j)}},
\label{facest}
\end{equation}
where $i,j$ run over all sites. We evaluate ${\cal S}({\bf k})$ for ${\bf k} \in [0,2\pi)\otimes [0,2\pi)$, 
since all the momenta that belong to the first Brillouin zone of each lattice have an equivalent point 
in this region; for reference, the edges of the Brillouin zones will be displayed in the figures. The 
$k_y$ component of each momentum is unequivocally determined by the periodic boundary conditions along 
the $y$ direction; on the other hand, the open boundary conditions along the $x$ direction do not 
impose any restriction for the $k_x$ component. Therefore, we can take advantage of this freedom to 
circumvent the discreteness of $\bf k$ in the $x$ direction.
\begin{figure}[h]
\centering
\includegraphics[width=0.45\textwidth]{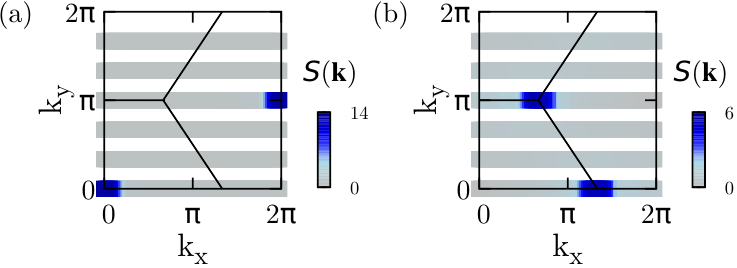}
\caption{Intensity plot of ${\cal S}(\bf k)$ for the triangular lattice with (a) $t'=0$, and 
(b) $t'=t$. The solid lines indicate the edges of the hexagonal Brillouin zones. Notice the 
discreteness of $k_y$ values.}
\label{orden-t}
\end{figure}

We have observed that ${\cal S}(\bf k)$ exhibits a pronounced peak for a certain momentum (and 
equivalent points in the reciprocal space), for both lattices, and all $t'$, except very close to the 
critical $t'_c,$ as we will discuss later. The intensity of the peak increases linearly with the 
cluster size, pointing out the existence of long-range magnetic order, and its position determines 
the magnetic wave vector ${\bf Q}$ of the magnetic order. Therefore, besides the Nagaoka state, 
long-range magnetic order is ubiquitous for the studied systems. Next, we analyze the magnetic pattern 
as a function of the kinetic energy frustrating hoppings $t'$. 

First, we present the results for the triangular lattice. We have chosen the $N=54$ sites cluster for 
the presentation of ${\cal S}(\bf k)$ as, for $L_y=6$, it is expected to be the cluster most 
representative of the two-dimensional case \cite{white07}. We begin revisiting the two previously 
known situations: $t'=0$ and $t'=t$. For $t'=0,$ Nagaoka's theorem is valid and, consequently, 
${\cal S}(\bf k)$ exhibits a sharp maximum at the magnetic wave vector ${\bf Q}={\bf 0}$, as it is 
shown in Fig. \ref{orden-t}(a). The other known case, $t'=t$ \cite{sposetti14}, is presented in 
Fig. \ref{orden-t}(b). We can see two ${\cal S}(\bf k)$ maxima at 
${\bf Q}\!\!=\!\!\left({\frac{4\pi}{3}},0\right)$ and 
${\bf Q^*}\!\!=\!\!\left({\frac{2\pi}{3}},{\frac{2\pi}{\sqrt{3}}}\right)$, both equivalent and 
corresponding to  the $120^\circ$ N\'eel order. 

\begin{figure}[h]
\centering
\includegraphics[width=0.40\textwidth]{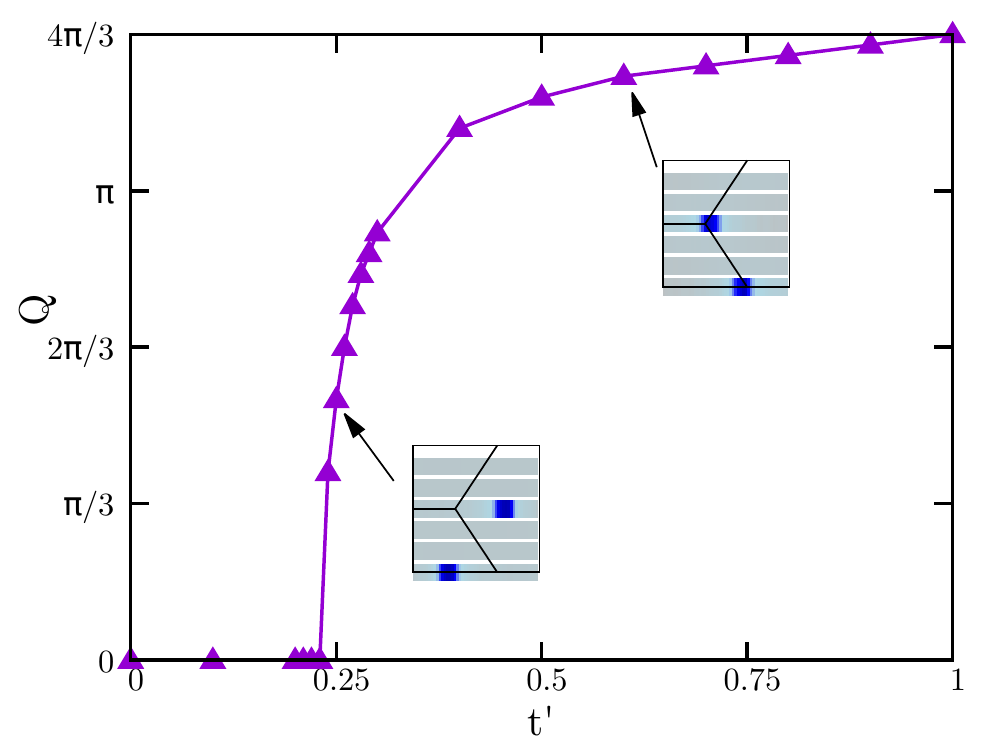}
\caption{Spiral pitch $Q$ as a function of $t'$ for the triangular lattice. Insets: intensity 
plots of ${\cal S}(\bf k)$ for $t' = 0.25$ and $0.6$. The darker regions correspond to the 
magnetic wave vector ${\bf Q}=(Q,0)$ (and other equivalent point).} 
\label{qtp}
\end{figure}

Now, we analyze the magnetic order for intermediate $t'$ between the two limits presented 
above. As long as $t' < t'^{\triangle}_c,$ we have found ferromagnetic order, that is 
${\bf Q}=\bf 0,$ in agreement with the ground state energy analysis of the previous section.  
Increasing $t'$ beyond the critical value, the ${\cal S}(\bf k)$ peak locates at ${\bf Q}=(Q,0)$, 
with $Q$ changing monotonously from $Q=0$ at $t'=t'^{\triangle}_c$ to $Q=\frac{4\pi}{3}$ at
$t'= t$, as it is shown in Fig. \ref{qtp}. 
This magnetic wave vector is characteristic of a 
spiral order with a pitch angle $Q$ in the $x$ direction, and it connects the ferromagnetic 
and the $120^{\circ}$ N\'eel orders. From the figure, it seems that $Q$ depends continuously 
on $t'$, although it exhibits a sharp rise close to $t'_c$. For small spiral pitch, the 
period of the spin pattern along the $x$ direction ($\sim \frac{2\pi}{Q}$) can exceed the 
cluster length $L_y$ ($Q \lesssim \frac{\pi}{4}$ in our case), preventing a correct 
description of the spiral order. A clear manifestation of this kind of finite size effects is 
that, very close to the transition, the peaks of the magnetic structure factor are not so 
pronounced. For these reasons, we can not state categorically that the transition from 
the Nagaoka state to the spiral one is continuous, although the $Q$ dependence on $t'$ 
suggest it. To underline this point, it is worth remembering that, in the previous section, 
we have seen the ground state energy for the triangular lattice also behaves rather smooth 
across the transition. We can speculate that an infinite-order phase transition takes place
here, like the one found as a function of doping, for $t'=0$, in Ref. \cite{carleo11}.

From Fig. \ref{qtp} we notice that, due to the sharp rise of $Q$, the spiral order has a 
pervasive antiferromagnetic character ($Q \gtrsim \pi$), except near the transition 
point. 

\begin{figure}[h]
\centering
\includegraphics[width=0.45\textwidth]{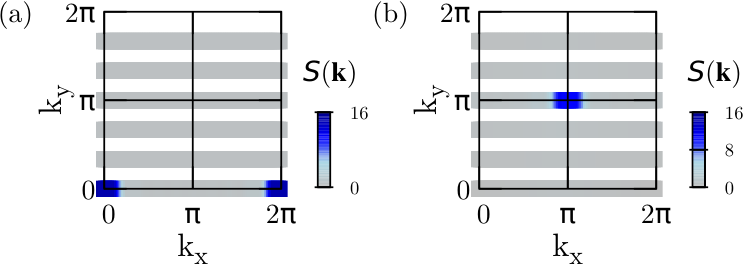}
\caption{Intensity plot of ${\cal S}(\bf k)$ for the square lattice with (a) $t'=0$, and (b) $t'=t$. 
Solid lines refer to the edges of the square Brillouin zones.}
\label{orden-c}
\end{figure}

In the following, we focus on the square lattice model. In Figs. \ref{orden-c}(a) and \ref{orden-c}(b)
we show the $t'=0$ and $t'=t$ intensity plot of ${\cal S}(\bf k)$, respectively. We have chosen the $N=60$ 
cluster in this case \cite{white07}. We can see in Fig. \ref{orden-c}(a) that the ground state has ferromagnetic 
order, as it was expected, because of the validity of Nagaoka's theorem for $t'=0$. For $t'=t$, 
Fig. \ref{orden-c}(b) indicates that the ground state has the typical N\'eel order with magnetic wave vector 
${\bf Q}\!=\!\left(\pi,\pi\right),$ as was found in Ref. \cite{sposetti14}.
The magnetic order evolution with second-nearest neighbor hopping $t'$ is very simple. 
Below $t'^{\square}_{c}$ the ground state is the Nagaoka ferromagnet, while for $t' > t'^{\square}_c$ 
it is the ${\bf Q}=(\pi,\pi)$ N\'eel order. Therefore, the transition is clearly discontinuous, with 
no intermediate order between the ${\bf Q}=(0,0)$ Nagaoka and ${\bf Q}=(\pi,\pi)$ N\'eel states.
It should be remember that, in this case, the ground state energy (Fig. \ref{ene-c}) exhibits an 
appreciable non-analytical behavior at the transition. Similar discontinuous transitions occur 
when kinetic frustration is due to staggered magnetic fluxes \cite{wang08}.
Summarizing, the addition of rather small $t'$ hoppings ($t'/t \gtrsim 0.22$) destabilizes 
the Nagaoka state and it induces long-range antiferromagnetic order, whose characteristics depend 
on the system: a spiral 
pattern in the triangular lattice, and the usual N\'eel order in the square case. Concerning 
the physical origin of this kinetic antiferromagnetism, recently we identified its 
microscopic mechanism \cite{sposetti14}. While the introduction of $t'$  increases the hole 
kinetic energy (the only one involved for $U=\infty$) in a ferromagnetic background, 
quantum interference effects can release this kinetic energy frustration if the hole moves along a certain 
antiferromagnetic pattern. This release can occur in two different ways, depending on the lattice 
geometry and the hopping terms: (a) the hole acquires a non-trivial spin Berry phase due to the 
antiferromagnetic texture or, (b) the magnetic order leads to an 
effective vanishing of the hopping amplitude along the frustrating loops.
In our work, the spin-Berry phase mechanism is operating for the stabilization of the spiral 
order in the triangular lattice model since, in that order, the hole acquires a $\pi$ phase 
when it hopes along one elementary triangle. On the other hand, for the square case, the effective hole 
nearest-neighbor hopping amplitude vanishes as a consequence of the antiparallel spin structure of the 
$(\pi,\pi)$ N\'eel order, and the hole moves only along the diagonal directions. 
We refer the reader to Ref. \cite{sposetti14} (especially to the Supplemental 
Material), for a more detailed discussion of the kinetic antiferromagnetism mechanism.  

\subsection{Local magnetization}

Finally, we have calculated the magnetic order parameter, that is, the local staggered 
magnetization $M_s$, which can be defined as 
$$M_s = \sqrt{\frac{1}{N^2}\sum_\alpha\langle {\left(\sum_{i \in \alpha}{\bf S}_i\right)^2\rangle}},$$
where $\alpha$ denotes the magnetic sublattices. A straightforward calculation shows that 
$M_s = \sqrt{\frac{{\cal S}(\bf Q)}{N}}$. 
On the other hand, for non-collinear magnetic structures, like the ${\bf Q}=(Q,0)$ spiral, 
we use the relation $M_s = \sqrt{\frac{2{\cal S}(\bf Q)}{N}},$ that can be deduced considering
semi-classical spin correlations, $\langle {\bf S}_i \cdot {\bf S}_j\rangle \simeq M_s^2 
\cos {\bf Q}\cdot \left({\bf R}_j-{\bf R}_j\right),$ for $i \neq j$. The same procedure recovers 
$M_s = \sqrt{\frac{{\cal S}({\bf Q})}{N}}$ for collinear orders.
\begin{figure}[t]
\centering
\includegraphics[width=0.45\textwidth]{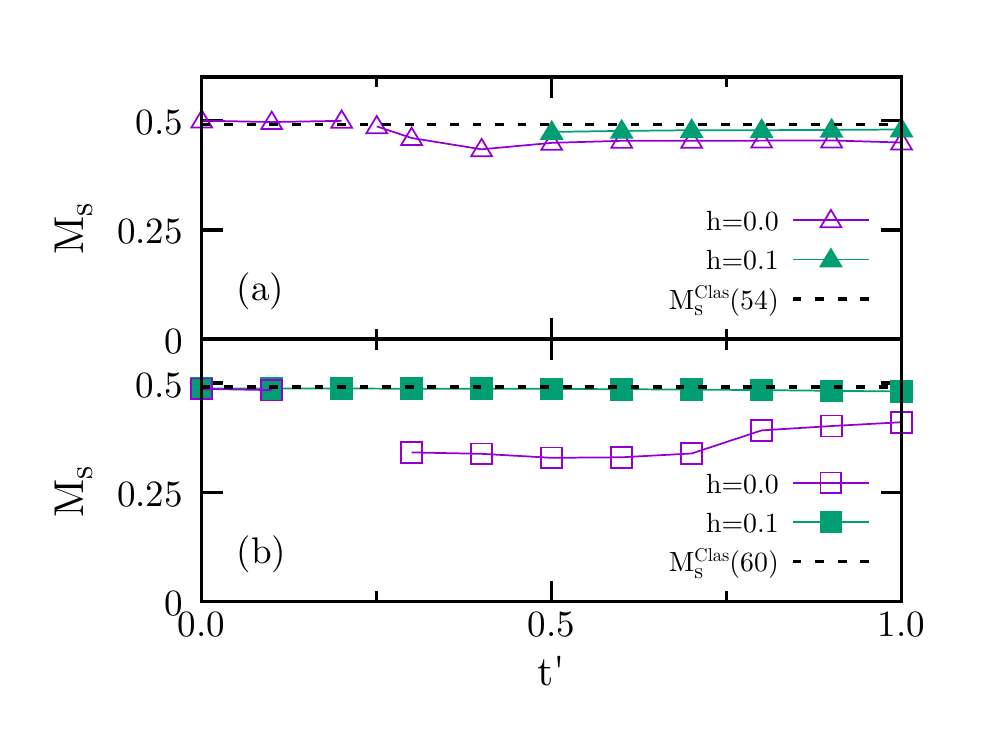}
\caption{Local magnetization $M_s$ as a function of $t'$ for (a) the $N=54$ triangular lattice 
without a magnetic field (open triangles) and with a magnetic field $h=0.1$ applied to one 
$120^{\circ}$ sublattice (solid triangles); (b) the $N=60$ square lattice without a magnetic field 
(open squares) and with a magnetic field $h=0.1$ applied to one of the two sublattices of the 
$(\pi,\pi)$ N\'eel order. The dashed lines correspond to the classical local magnetization, 
$M_s^{clas}=\frac{1}{2}-\frac{1}{2N}$.}
\label{ordenqx-t}
\end{figure}
The curve with open triangles (squares) in Fig. \ref{ordenqx-t} shows the evolution of the 
local magnetization with $t'$, for the $N=54$ ($N=60$) triangular (square) cluster. 
The behavior for other clusters is qualitatively similar. 
In the Nagaoka phase, $M_s$ takes the classical value, given by 
$ M_s^{clas} = \frac{1}{2}-\frac{1}{2N}$, as it was expected for a fully saturated state. 
On the other hand, the kinetic antiferromagnetic phases have large values of $M_s$, especially 
the triangular model (upper panel of the figure) where it is close to $M_s^{clas}$ for 
all $t'$. 

In Ref. \cite{sposetti14}, we found that, for the particular value $t'=t$ and both models, 
$M_s$ extrapolates to its classical value in the thermodynamic limit. There, we argued that 
the classical value is not reached for finite clusters due to the $SU(2)$ symmetry of the Hubbard model. 
One way to confirm this idea, for commensurate magnetic orders, was the finding that the application of 
a small uniform magnetic field ${\vec h}=h \hat{e}_z,$ in one sublattice only, was enough to pin the 
classical order for the finite clusters \cite{white07}. 

In this work, we wonder if the remarkable classical character mentioned above extends to other values 
of $t'$. Hence, we have applied a magnetic field $h = 0.1$ in one of the two (three) 
sublattices of the $180^{\circ}$ ($120^{\circ}$) structure in the square (triangular) lattice model.
Notice that, in the triangular case, due to the incommensuration of the ${\bf Q} = (Q,0)$ spiral phase for 
a generic $Q$, the application of the pinning magnetic field matches the order only when 
$Q \simeq \frac{4\pi}{3}$ (so, for $t'/t \lesssim 0.5$ we do not apply $h$). The curves with solid symbols in Fig. \ref{ordenqx-t} 
show the $t'$ dependence of $M_s$ with the magnetic field $h$ applied. Convincingly, it can be seen that 
$h$ brings out the ``hidden'' classical nature of the kinetic antiferromagnetism in these finite clusters.  
Therefore, we have found that, for all $t'$, the ground state of the considered $U=\infty$ Hubbard models 
has classical magnetism. Above $t'_c$, this is a remarkable result as it can be thought as a perfect 
counterpart of Nagaoka's theorem. About the physical reason for the classical nature of the 
antiferromagnetism, we speculate that the hole motion under the $U=\infty$ condition generates 
effective long-range spin interactions, that may favor classical ordering \cite{sposetti14,lieb62}.

\section{Conclusion}

To conclude, we have investigated the evolution of the Nagaoka state with $t'$ hopping processes 
that induce hole kinetic energy frustration. To this purpose, the numerical density matrix 
renormalization group is used to compute the magnetic ground state properties of the $U=\infty$ 
Hubbard model, with one hole doped away from half-filling, on two lattices: the spatially anisotropic triangular 
lattice and the square lattice with second-nearest neighbor hoppings. 

We have found that the Nagaoka ferromagnetic state is destabilized for rather small frustrating 
hoppings: $t'_{c}/t=0.222$ (0.214) for the triangular (square) lattice. Taking into account that these 
values are comparable to the corresponding ones for a simple 4-site cluster, we can state that 
Nagaoka physics is driven mainly by quantum interferences generated by the hole motion along the 
smallest lattice loops.

The analysis of the magnetic structure factor indicates that the ground state of the $U=\infty$ Hubbard 
model has long-range magnetic order for all hopping $t'$. For $t' > t'_c,$ the square lattice ground state 
has a classical $(\pi,\pi)$ N\'eel order. In the triangular case, the magnetic order
is a classical spiral pattern with magnetic wave vector ${\bf Q}=(Q,0)$, that seems to connect continuously 
the Nagaoka ferromagnet at $t'=t'_c$ with the $120^{\circ}$ N\'eel antiferromagnet at the isotropic point 
$t'=t$. 

Therefore, the kinetic antiferromagnetism, first discovered by Haerter and Shastry \cite{haerter05} on the
triangular lattice and further developed by us \cite{sposetti14}, seems to be a robust phenomenon in
kinetically frustrated electronic systems. Its classical nature makes this 
antiferromagnetism the perfect counterpart of the Nagaoka's state for frustrated systems. 
On the other hand, the lattice geometry affects the nature of the transition between the different 
classical states, as we have seen: continuous for the triangular lattice, discontinuous for the square one.

Finally, it remains a challenge to find experimental realizations of both, the old Nagaoka ferromagnetism and 
the new kinetic antiferromagnetism that have been studied in this and previous works \cite{sposetti14}. Up to date, there is no clear evidence of 
Nagaoka physics in condensed matter systems, being the main obstacle the large onsite Coulomb repulsion needed, 
$U/t \gtrsim 100$. However, tunable Feshbach resonances in ultracold atoms allow to reach such large $U'$s, and, 
on the other hand, the generation of artificial gauge fields in triangular optical lattices \cite{struck11} can 
induce frustrated motion. Other proposals involve artificial lattice of quantum dots \cite{gaudreau06}, 
and high-density two-dimensional electron gas at the interface between Mott and band insulators \cite{iaconis16}. 
All this opens the interesting possibility to test experimentally our proposals about kinetic magnetism. 

\acknowledgments 
This work was supported by CONICET-PIP1060, and from CONICET-PIP0389.


\begin{thebibliography}{99}

\bibitem{nagaoka66} Y. Nagaoka, 
Phys. Rev. {\bf 147}, 392 (1966).

\bibitem{hartmann93} E. M\"uller-Hartmann, Th. Hanisch, and R. Hirsch, 
Physica B {\bf 186-188}, 834 (1993).

\bibitem{fazekas99} P. Fazekas, 
{\it Electron Correlation and Magnetism} (World Scientific, Singapore, 1999).

\bibitem{park08} H. Park, K. Haule, C. A. Marianetti, and G. Kotliar, 
Phys. Rev. B, {\bf 77}, 035107 (2008).

\bibitem{liu12} L. Liu, H. Yao, E. Berg, S. R. White, and S. A. Kivelson, 
Phys. Rev. Lett. {\bf 108}, 126406 (2012).

\bibitem{liang95} S. Liang and H. Pang, 
Europhys. Lett. {\bf 32}, 173 (1995). 

\bibitem{carleo11} G. Carleo, S. Moroni, F. Becca, and S. Baroni, 
Phys. Rev. B, {\bf 83}, 060411 R, (2011)

\bibitem{eisenberg02} E. Eisenberg, R. Berkovits, D. A. Huse, and B. L. Altshuler, 
Phys. Rev. B {\bf 65}, 134437 (2002). 

\bibitem{oles91} A. M. Ol\'es and P. Prelovsek, Phys. Rev. B {\bf 43}, 13348 (1991). 


\bibitem{barford91} W. Barford and J. H. Kim, 
Phys. Rev. B {\bf 43}, 559 (1991).

\bibitem{merino06} J. Merino, B. J. Powell, and R. H. McKenzie, 
Phys. Rev. B {\bf 73}, 235107 (2006). 

\bibitem{wang08} Y. F. Wang, C.D. Gong, and Z. D. Wang, 
Phys. Rev. Lett. {\bf 100}, 037202 (2008).

\bibitem{haerter05} J. O. Haerter and B. S. Shastry, Phys. Rev. Lett. {\bf 95},087202 (2005).

\bibitem{sposetti14} C. N. Sposetti, B. Bravo, A. E. Trumper, C. J. Gazza, and L. O. Manuel, 
Phys. Rev. Lett. {\bf 112}, 187204 (2014).

\bibitem{white92} S. R. White, 
Phys. Rev. Lett. {\bf 69}, 2863, (1992).

\bibitem{white93} S. R. White and D. A. Huse, 
Phys. Rev. B {\bf 48}, 3844 (1993).

\bibitem{brinkman70} W. F. Brinkman and T. M. Rice, 
Phys. Rev. B {\bf 2}, 1324 (1970).

\bibitem{white07} S. R. White and A. L. Chernyshev, 
Phys. Rev. Lett. {\bf 99}, 127004 (2007).

\bibitem{lieb62} E. Lieb and D. Mattis,
J. Math. Phys. {\bf 3}, 749 (1962).

\bibitem{struck11} J. Struck, C. \"Olschl\"ager, R.L. Targat, P. Soltan-Panahi, A. Eckardt, M. Lewenstein,  P. Windpassinger, and K. Sengstock, 
Science {\bf 333}, 996 (2011). 

\bibitem{gaudreau06} L. Gaudreau, S. A. Studenikin, A. S. Sachrajda, P. Zawadzki, A. Kam, J. Lapointe, M. Korkusinski, and P. Hawrylak, 
Phys. Rev. Lett. {\bf 97}, 036807 (2006).

\bibitem{iaconis16} J. Iaconis, H. Ishizuka, D. N. Sheng, and L. Balents, 
Phys. Rev. B {\bf 93}, 155144 (2016). 

\end{thebibliography}
\end{document}